%% file: main.tex
\documentclass[conference]{IEEEtran}
\IEEEoverridecommandlockouts
\usepackage{cite}
\usepackage{hyperref}
\usepackage{amsmath,amssymb,amsfonts}
\usepackage{bbm}
\usepackage{algorithm}
\usepackage{algpseudocode}
\usepackage{graphicx}
\usepackage{textcomp}
\usepackage{xcolor}
\def\BibTeX{{\rm B\kern-.05em{\sc i\kern-.025em b}\kern-.08em
    T\kern-.1667em\lower.7ex\hbox{E}\kern-.125emX}}
\usepackage{mathtools}
\usepackage{booktabs}
\usepackage{multirow}
\usepackage{subfig}
\IEEEsettopmargin{t}{0.8in}

\newcommand{\E}[1]{\mathbb{E}\left[ #1 \right]} 
\newcommand{\mc}[1]{\mathcal{#1}}   
\newcommand{\mb}[1]{\mathbf{#1}}    
\DeclareMathOperator*{\argmax}{arg\,max}    

\usepackage{glossaries}
\newacronym{drl}{DRL}{Deep Reinforcement Learning}
\newacronym{qos}{QoS}{Quality of Service}
\newacronym{kpi}{KPI}{Key Performance Indicator}
\newacronym{iot}{IoT}{Internet of Things}
\newacronym{ue}{UE}{User Equipment}
\newacronym{mdp}{MDP}{Markov Decision Process}
\newacronym{fifo}{FIFO}{First-In First-Out}
\newacronym{m2m}{M2M}{Machine to Machine}
\newacronym{urllc}{URLLC}{Ultra-Reliable Low-Latency Communications}
\newacronym{embb}{eMBB}{enhanced Mobile Broadband}
\newacronym{ofdma}{OFDMA}{Orthogonal Frequency Division Multiple Access}
\newacronym{vr}{VR}{Virtual Reality}
\newacronym{mec}{MEC}{Mobile Edge Computing}
\newacronym{ai}{AI}{Artificial Intelligence}
\newacronym{fl}{FL}{Federated Learning}
\newacronym{dqn}{DQN}{Deep Q-Network}
\newacronym{cdf}{CDF}{Cumulative Distribution Function}

\definecolor{color0}{HTML}{0000FF}
\definecolor{color1}{HTML}{B13AC5}
\definecolor{color2}{HTML}{E6748D}
\definecolor{color3}{HTML}{FFB14E}

\def \fwidth{0.95\columnwidth}
\def \fheight {0.5\columnwidth}

\newcommand\copyrighttext{%
  \footnotesize \textcopyright 2022 IEEE. This paper has been submitted to IEEE ICC 2023. Personal use of this material is permitted. Permission from IEEE must be obtained for all
other uses, in any current or future media, including reprinting/republishing this material for advertising or promotional purposes, creating new collective
works, for resale or redistribution to servers or lists, or reuse of any copyrighted component of this work in other works}
\newcommand\copyrightnotice{%
\begin{tikzpicture}[remember picture,overlay]
\node[anchor=south,yshift=10pt] at (current page.south) {\fbox{\parbox{\dimexpr\textwidth-\fboxsep-\fboxrule\relax}{\copyrighttext}}};
\end{tikzpicture}%
}

\usepackage[utf8]{inputenc}
\usepackage{pgfplots}
\DeclareUnicodeCharacter{2212}{−}
\usepgfplotslibrary{groupplots,dateplot}
\usetikzlibrary{patterns,shapes.arrows}
\pgfplotsset{compat=newest}
\begin{document}

\title{
\copyrightnotice
The Cost of Learning: Efficiency \textit{vs.} Efficacy of Learning-Based RRM for 6G
\thanks{This work was supported by the EU H2020 MSCA ITN project Greenedge (grant no. 953775), by the MUR (Italian Ministry for Universities and Research) under the PNRR PE 14 project "Restart", and by the Villum Foundation, Denmark, under Villum Investigator grant ``WATER.''}
}

\author{\IEEEauthorblockN{Seyyidahmed Lahmer\IEEEauthorrefmark{1}, Federico Chiariotti\IEEEauthorrefmark{1}\IEEEauthorrefmark{2}, Andrea Zanella\IEEEauthorrefmark{1}}
\IEEEauthorblockA{\IEEEauthorrefmark{1}Department of Information Engineering, University of Padova, via G. Gradenigo 6B, 35131 Padova, Italy\\
\IEEEauthorrefmark{2}Department of Electronic Systems, Aalborg University, Fredrik Bajers Vej 7C, 9220 Aalborg \O{}st, Denmark\\
Emails: {\texttt{\{federico.chiariotti, seyyidahmed.lahmer, andrea.zanella\}@unipd.it}}}
}

\newcommand{\AZ}[1]{\textcolor{blue}{AZ: #1}}

\maketitle

\begin{abstract}
In the past few years, \gls{drl} has become a valuable solution to automatically learn efficient resource management strategies in complex networks. In many scenarios, the learning task is performed in the Cloud, while experience samples are generated directly by edge nodes or users. Therefore, the learning task involves some data exchange which, in turn, subtracts a certain amount of transmission resources from the system. This creates a friction between the need to speed up convergence towards an \textit{effective} strategy, which requires the allocation of resources to transmit learning samples, and the need to maximize the amount of resources used for data plane communication, maximizing users' \gls{qos}, which requires the learning process to be \textit{efficient}, i.e., minimize its overhead.
In this paper, we investigate this trade-off and propose a dynamic balancing strategy between the learning and data planes, which allows the centralized learning agent to quickly converge to an efficient resource allocation strategy, while minimizing the impact on \gls{qos}. Simulation results show that the proposed method outperforms static allocation methods, converging to the optimal policy (i.e., maximum efficacy and minimum overhead of the learning plane) in the long run.
\end{abstract}


\section{Introduction}\label{sec:intro}

The use of \gls{ai} in communication networks has become pervasive with the transition from 4G to 5G, and learning is at the core of the 6G standardization process~\cite{letaief2019roadmap}: mobile networks have become exponentially more complex, with multiple \gls{qos} targets and extremely fast dynamics, and computational and energetic considerations have become inextricable from communications with the rise of green networking~\cite{chih2021ai} and \gls{mec}~\cite{hu2021energy}. The 6G vision acknowledges that hand-designed resource allocation strategies are not up to the challenge of managing all these elements, proposing the use of \gls{drl} as an adaptable and robust alternative for network orchestration~\cite{rlInNetworkingSurvey} and resource allocation~\cite{sami2021ai}, along with a variety of other optimization tasks. \gls{drl}'s \emph{effectiveness} in dealing with complex scenarios is well-established: these agents can find foresighted policies aiming for long-term objectives~\cite{sutton2018reinforcement}, significantly improving network performance after the \gls{drl} agent has been trained.

However, large and complex \gls{drl} models, such as those required to control modern communication networks, are also computational processes that require non-negligible computational~\cite{neda2022survey}, transmission~\cite{beaumont2021efficient}, and energetic resources. Local training at the edge puts a significant strain on \gls{mec} nodes with limited energetic and computational budgets, and offloading training to the Cloud incurs a significant communication overhead.
As the use of \gls{drl} in 6G is aimed at optimizing the allocation of communication and computational resources, not considering the \emph{cost of learning} might lead to suboptimal outcomes~\cite{jang2020knowledge}. In this sense, their \emph{efficiency} becomes questionable: the performance during the training process might be affected by the cost of the process itself, particularly for more complex models and agents, leading to a trade-off between it and effectiveness at convergence.

Online training, which is necessary in environments with time-varying statistics, is particularly vulnerable to this issue~\cite{villacca2021online}, as the \gls{drl} agent needs to keep training to update its policy to the changes in the environment. In our previous work~\cite{mason2022no}, we considered the trade-off between effectiveness and efficiency of Cloud-based \gls{drl} training by optimizing the resource allocation between the data plane, i.e., the network resources allocated to the end users, and the \emph{learning plane}, i.e., the portion of the network resources used by the learning process. However, 
in that work, we considered a static resource allocation to the learning plane, thus fixing the overhead cost of the learning process on the system and, at the same time, fixing the speed at which the \gls{drl} could converge towards an efficient strategy for the allocation of the remaining resources to the users. 

In this work, we go beyond a simple static allocation and define a more general learning plane control loop, which decides the allocation of resources based on a greedy optimization strategy. This dynamic resource allocation is computationally simple, and maintains a similar efficiency to fixed allocation strategies, while reaching the same effectiveness as ideal out-of-band (i.e., with negligible overhead) approaches after the training period is over. We demonstrate the performance of the proposed framework in a simple network slicing scenario, showing that the dynamic approach leads to significantly better performance after an initial transition. To be noticed that the optimization framework we propose is not tied to the networking scenario, but works in any allocation problem in which the pool of resources that the \gls{drl} agent needs to allocate to the users is also required for the training of the \gls{drl} agent itself.

The rest of this paper is divided as follows: first, Sec.~\ref{sec:system} presents the system model and the learning plane optimization. The slicing use case is then described in Sec.~\ref{sec:slicing}, while Sec.~\ref{sec:results} presents the simulation results. Finally, Sec.~\ref{sec:conc} concludes the paper and presents some possible future work on the subject.


\section{System Model}\label{sec:system}
\begin{figure*}[!t]
\centering
    \includegraphics[width=0.9\textwidth]{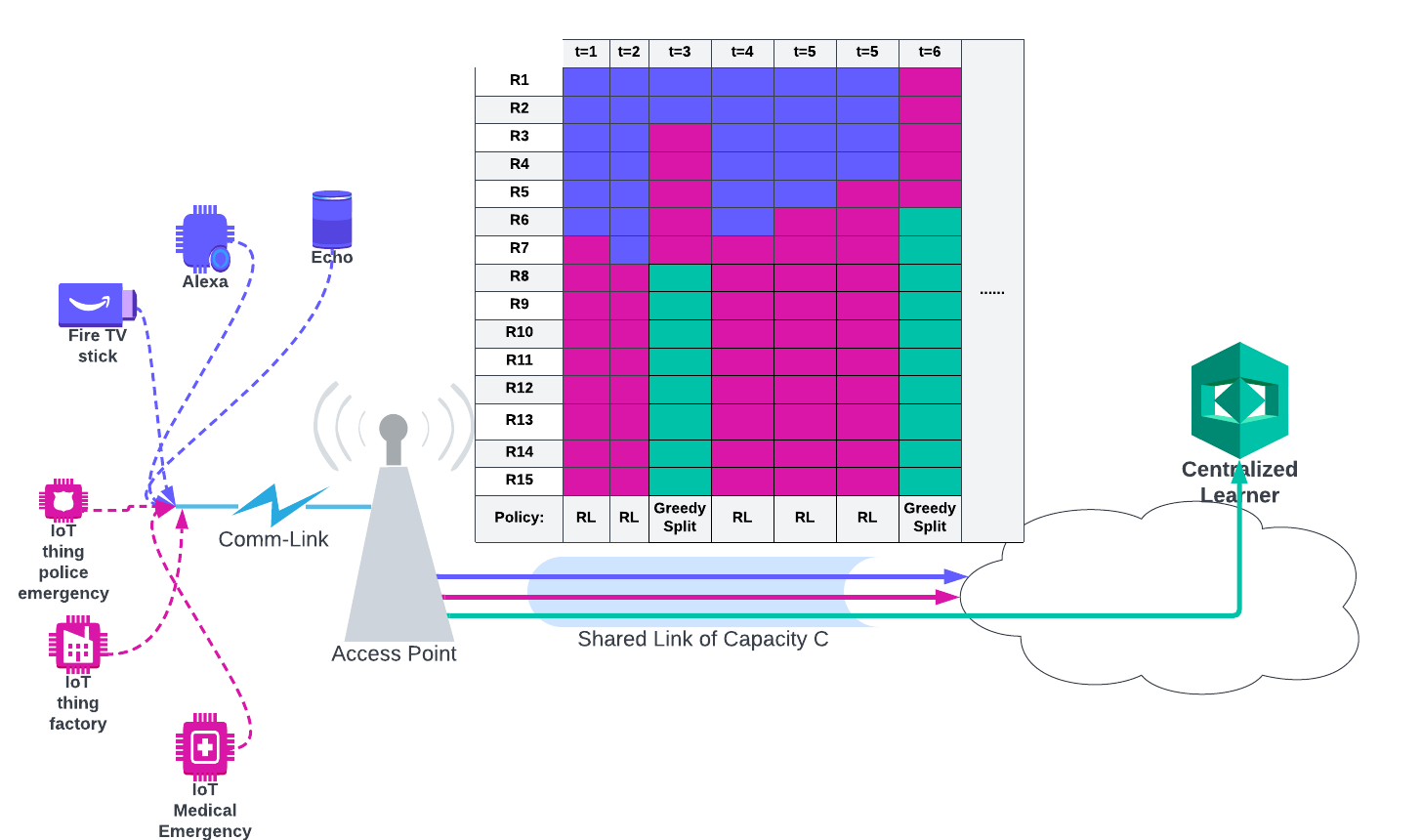}
    \label{fig:scenario}
    \caption{Schematic of the learning control loop in a communication scenario.}
    \label{fig:schematic}
\end{figure*}

Let us consider a generic resource allocation problem, which is modeled as an infinite horizon \gls{mdp} defined by the tuple $(\mc{S},\mc{A},\mb{P},R,\gamma)$: $\mc{S}$ represents the state space, $\mc{A}$ is the action space (which is potentially different for each state), $\mb{P}\in[0,1]^{|\mc{S}|\times|\mc{A}|\times|\mc{S}|}$ is the transition probability matrix, which considers both the state and the action chosen by the agent, $R:\mc{S}\times\mc{A}\times\mc{S}\rightarrow\mathbb{R}$ is the reward function, and $\gamma\in[0,1)$ is the discount factor. The ultimate objective of a \gls{drl} agent is to find the optimal policy $\pi^*:\mc{S}\rightarrow\mc{A}$, which maximizes the expected long-term reward:
\begin{equation}
    \pi^*=\argmax_{\pi:\mc{S}\rightarrow\mc{A}}\E{\sum_{t=0}^{\infty}\gamma^t R(\mb{s}_t,\pi(\mb{s}_t),\mb{s}_{t+1})}.
\end{equation}

Time is divided in slots, and the slot index is denoted by $t \in \mathbb{Z^{+}}$. Let us assume that, in each time slot $t$ and independently from the state $s\in\mc{S}$, the system can allocate $N$ resource blocks, which may represent communication bandwidth, computational cycles, or energy, depending on the specific application: the type of resource may affect the definition of the specific \gls{mdp}, but is irrelevant at this point. In general, we refer to \emph{requests} in the following: a request may be a packet that needs to be transmitted, a computing job that needs to be executed, or an action that requires some energy, but each request requires exactly one resource block. 

The system resources are allocated among $M$ different \textit{slices}, where a slice may represent a single user, or a group of users with the same features. The action space then contains all possible resource allocation vectors that split the $N$ resources among the $M$ slices:
\begin{equation}
    \mc{A}=\left\{\mb{a}\in\{0,\ldots,N\}^M:\sum_{m=1}^M a_m= N\right\}.\label{eq:general_action_space}
\end{equation}

Furthermore, we assume that each slice is associated to a \gls{fifo} queue of requests: each queue has a limited size $Q$, after which the system starts dropping older requests for that slice to make room for newer ones. 

In this work, we focus on \glspl{kpi} tied to the latency with which the requests of the different slices are served. However, the approach can be generalized to consider other metrics. 

We hence indicate by $T_{m,i}$ the latency of the $i$-th request from slice $m$, which depends on the time it spends in the queue before being assigned a resource. Dropped or rejected requests have an infinite latency by definition. The $i$-th request from slice $m$ is generated at time $t_{m,i}$, and age $\Delta_{m,i}(t)$ is defined as:
\begin{equation}
    \Delta_{m,i}(t)=t-t_{m,i}.
\end{equation}
We can then define the reward function:
\begin{equation}
    R(\mb{s},\mb{a},\mb{s}')=\sum_{m=1}^M\sum_{i=1}^{a_m}f_m\left(\Delta_{q_m(i)}\right),\label{eq:general_reward}
\end{equation}
where $\Delta_{q_m(i)}$ is the age of the packet in position $i$ of the $m$-th queue at the current time $t$, and $f_m:\mathbb{N}\rightarrow[0,1]$ is a function mapping the latency of each request to slice $m$'s resulting \gls{qos}. With a slight abuse of notation, we define $f(\varnothing)=0$, where $\varnothing$ indicates that there is no packet in that position in the queue. We can distinguish between slices with \emph{hard} timing requirements, for which the \gls{qos} of a request is 1 if it is served within a maximum latency, and 0 if it exceeds that deadline; and \emph{soft} timing requirements, for which the \gls{qos} is a generic monotonically decreasing function of the latency. It should also be noted that dropped or rejected requests do not generate any rewards, as they are never included in the sum. The state of the system is then represented by the age of each request contained in each queue, so that in the most general case, $\mc{S}=\left(\{\varnothing\}\cup\mathbb{N}\right)^{M\times Q}$.

The objective of the learning agent is then to learn how to allocate resources among users, so as to maximize their \gls{qos} parameters; it should also be aware of the slices that have a higher risk of violating hard timing requirements and schedule resources to avoid missing deadlines. However, learning is also a computational process, and the \gls{drl} agent may take up some of the same resources that may be allocated to the users in order to improve its policy. As we highlighted in our previous work~\cite{mason2022no}, considering the cost of learning can lead to significantly different choices, limiting the amount and type of experience samples that are selected for training: this is also true regardless of the type of resource the learning requires.

However, even our previous work only considered static policies, which set up a separate virtual channel (either divided in time or in frequency) for the learning data, strictly separating the learning and data planes. Equivalently, an agent learning how to schedule tasks in an edge server could reserve a certain percentage of computation time to self-improvement, but the amount was decided in advance. This is clearly suboptimal: intuitively, the relative returns from policy self-improvement decrease over time, as the agent gradually converges to the optimal policy. After convergence, and as long as the environment statistics are stable, the value of further improvements to the policy is zero by definition. A dynamic policy for adapting the allocation between requests and learning should then take this into account. 

Furthermore, the current state of the system also needs to be taken into account: if delaying the queued requests further does not have a large impact on the \gls{qos}, the system can take away resources from the slices in order to improve the resource allocation policy, but if the impact is big, e.g., if some requests from a slice with hard timing requirements are already close to the deadline, they need to be prioritized, choosing immediate gains over potential future improvements.

This is particularly important for non-stationary environments, in which the coherence time of the \gls{mdp} statistics is finite: in this kind of system, the learning agent needs to adapt the allocation to the changing statistics of the environment, and cannot rely on offline training, but must keep learning from experience and adapt to the changes proactively.

\subsection{Learning Plane Resource Allocation}

One of the problems of including the learning plane in the resource allocation policy is the circularity of the policy: in order to learn when to allocate resources to policy improvement, a \gls{drl} agent needs to first learn when learning is important. As the policy evolves over time, this makes the reward that the agent perceives dependent on the agent's own policy, making the environment non-stationary.

In order to avoid this problem, we need to set an external rule to regulate learning, so that the environment that the agent sees is stationary. We define a generic resource allocation vector space $\mc{Z}$ as follows:
\begin{equation}
    \mc{Z}=\left\{\mb{z}\in\{0,\ldots,N\}^M:\sum_{m=1}^M z_m\leq N\right\}.
\end{equation}
We remark here that $\mc{Z}$ is not the action space, but rather a superset of it, i.e., $\mc{A}\subseteq\mc{Z}$: the definition of the action space in~\eqref{eq:general_action_space} only considers allocation that gives all of the available resources to slices. If $\sum_{m=1}^M z_m<N$, the remaining resources are allocated to the learning.

We can then divide time slots in two categories: in \gls{drl} slots, the \gls{drl} agent applies its current policy and allocates all the resources to the slices requesting them, while in learning slots, we use a different policy, which divides resources between the learning process and the slices. Naturally, this policy must be simpler than the one defined by the \gls{drl} problem, and should not be learning-based, to avoid the circularity problem. We also remark that learning slots are not considered as experience samples for the \gls{drl} training, as the resource allocation $\mb{z}$ might not be a valid action in the \gls{mdp}.

Fig.~\ref{fig:schematic} shows a basic schematic of the process in the communication use case: the two classes of users, corresponding to \gls{iot} and human communications, transmit over a shared link, and the resources in each time slot (which correspond to bandwidth and time resources in the uplink to the Cloud) are allocated following the dynamic division. Slots 3 and 6 in the figure are learning slots, as a significant portion of the resources is allocated to the learning plane.

We can then define a randomized selection between \gls{drl} and learning slots: in each time slot $t$, the learning plane can be allocated some resources with probability $\rho(t)$, which decreases linearly over time, following a similar profile to the $\varepsilon$-greedy policy's exploration parameter. If we consider a coherence time for the scenario of $\tau$ slots, i.e., the statistics of the environment will be approximately stationary for $\tau$ steps, so that previous experience is still useful and the optimal policy is static, we can adapt the learning curve. However, we still need to define an allocation strategy in learning slots.

\subsection{Greedy Allocation Strategy}

\begin{figure}[t!]
\centering
\label{fig:allocationPolicy}
\centerline{\includegraphics[width=3.4in]{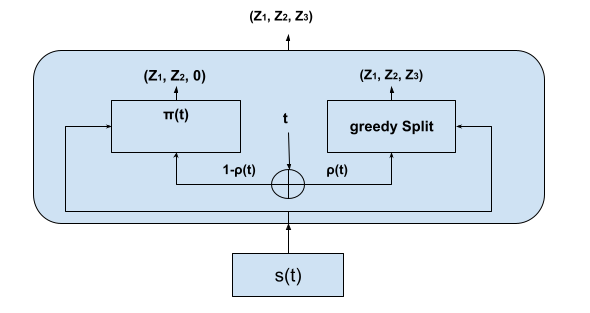}}
\caption{Schematic of the learning plane resource allocation policy.}
\label{fig:greedy}
\end{figure}

Firstly, we define a function $\hat{R}:\mc{S}\times\mc{Z}\rightarrow\mathbb{R}$, which represents the best approximation of the instantaneous reward for each resource allocation, considering only the information available in the current state. If the $f_m$ \gls{qos} functions are known, we can consider the following approximated reward:
\begin{equation}
    \hat{R}(\mb{s},\mb{z})=\sum_{m=1}^M\left[\sum_{i=1}^{z_m}f_m\left(\Delta_{q_m(i)}\right)-\sum_{\mathclap{j=z_m+1}}^{L}f_m\left(\Delta_{q_m(i)}+1\right)\right].
\end{equation}
Naturally, this only considers the instantaneous reward, and an allocation based on this function will often lead to worse outcomes: however, this is a simple policy that can be evaluated instantly, and its optimization is relatively simple.

The second element that we must take into account when designing the policy in learning slots: after each \gls{drl} slot, we generate an experience sample, which is queued for transmission or training. Due to memory limitations, the maximum number of unprocessed samples is $E$, and each sample is split into $\ell$ packets, each of which requires one resource allocated to the learning. We then define a second function $S(\mb{z},e)$, where $e\in\{0,\ldots,E\}$ is the number of samples in the experience queue. The greedy strategy is then the solution to the following optimization problem:
\begin{equation}
    \mb{z}^*(\mb{s},\mb{z},e)=\argmax_{\mb{z}}\left(M^{-1}\hat{R}(\mb{s},\mb{z})+E^{-1}S(\mb{z},e)\right).
\end{equation}
If $f_m$ is concave for all slices with a soft timing requirement, we can exploit the \gls{fifo} nature of the queue to provide a simple iterative solution, starting from the empty assignment and gradually assigning resources to either one of the slices or the learning process, depending on its value. 

Fig.~\ref{fig:greedy} shows a schematic of the full learning plane resource allocation strategy, in a simple case with $M=2$: at each time step, the node randomly selects either the \gls{drl} agent or (with probability $\rho(t)$) the greedy allocation, which reserves some resources for the learning plane, while taking care to avoid damaging the reward.

\section{Network Slicing Use Case}\label{sec:slicing}

\setlength{\textfloatsep}{15pt}
\begin{table}[t]
    \centering
    \caption{Use case and learning parameters.}
    \footnotesize
    {\renewcommand{\arraystretch}{1.25}
    \begin{tabular}{@{}rlcc@{}}
    \toprule
    \multicolumn{2}{c}{Parameter} & Symbol & Value \\ \midrule
    \multicolumn{4}{c}{\textbf{Communication system}}\\ \midrule
    \multicolumn{2}{c}{Number of subchannels} & $N$ & 15\\
    \multicolumn{2}{c}{Slot time duration} & $\tau$ & 1~ms\\
    \multicolumn{2}{c}{Packet queue length} & $Q$ & 1500 \\
    \multicolumn{2}{c}{Packet size} & $L$ & 512~B \\
    \multicolumn{2}{c}{Link capacity} & $C$ & 7.68~MB/s \\

     \midrule
    \multicolumn{4}{c}{\textbf{Traffic model}}\\ \midrule
     \multirow{4}{*}{Slice 1} & Total users & $U_1$ & 16\\     
     & Active user rate & $R_1$ & 512~kB/s\\
     & Activity transition matrix & $\mathbf{O}^{(1)}$ & $\begin{pmatrix} 0.5 & 0.5\\
     0.92 & 0.08\\
     \end{pmatrix}$\\
     & Total expected traffic & $\E{G_1}$ & 2.88~MB/s\\
     \midrule
     \multirow{5}{*}{Slice 2} & Total users & $U_2$ & 17\\
     & Active user rate & $R_2$ & 512~kB/s\\
     & Activity transition matrix & $\mathbf{O}^{(2)}$ & $\begin{pmatrix} 0.5 & 0.5\\
     0.5 & 0.5\\
     \end{pmatrix}$\\ 
     & Total expected traffic & $\E{G_2}$ & 4.35~MB/s\\
     & Packet deadline & $T_{\max}^{(2)}$ & 70~ms\\
     \midrule
     \multicolumn{4}{c}{\textbf{Learning plane}}\\ \midrule
     \multicolumn{2}{c}{Discount factor} & $\gamma$ & 0.95\\
     \multicolumn{2}{c}{Learning queue length} & $E$ & 1500\\
     \multicolumn{2}{c}{Packets required for each sample} & $\ell$ & 3\\
     \multicolumn{2}{c}{Initial learning slot probability} & $\rho_0$ & 0.2\\
     \multicolumn{2}{c}{Final learning slot probability} & $\rho_f$ & 0.01\\
     \multicolumn{2}{c}{Learning slot probability decay} & $\sigma$ & $8\times10^{-4}$\\
     \multicolumn{2}{c}{Learning slot decay pace} & $H$ & 1000\\
     \multicolumn{2}{c}{Queue pressure parameter} & $\chi_1$ & 1400\\

     \bottomrule
    \end{tabular}}
    \label{tab:params}
\end{table}

We consider a simple, but representative scenario, in which an \gls{ofdma} link with $N$ orthogonal subchannels is used to transmit the data packets generated by two classes of data sources, as well as traffic on the learning plane. The scenario fits the general model presented in the previous section, as the communication resources are shared between the data and learning planes. The full parameters for the scenario, which we will describe in this section, are given in Table~\ref{tab:params}.

\subsection{Communication System Model}
The number of active users in each slice is variable, making traffic non-deterministic. We consider two slices, corresponding to the two types of data sources:
\begin{itemize}
    \item Slice 1 represents file transfer traffic, for which we do not set any strict latency constraints. However, we want the system to have the highest possible reliability to ease the burden on the higher layers. As such, $f_1(T)=1$ for all finite values of $T$, but the \gls{qos} is 0 if $T$ is infinite (i.e., if the packet is dropped);
    \item Slice 2 represents interactive traffic, such as video conferencing or \gls{vr} traffic, with a strict latency deadline: packets need to be transmitted within a maximum latency $T_{\max}^{(2)}$, or the \gls{qos} drops to 0.
\end{itemize}

We consider a maximum number of active users $U_m\in\mathbb{N}$ for each slice, each of which follows a Gilbert-Elliott on-off model, which can be modeled as a binary Markov chain with transition probability matrix $\mb{O}^{(m)}$. In state 0, the user does not transmit, while in state 1, it transmits packets of size $L$ with a constant bitrate $R_m$.

The aggregate traffic generated by slice $m$ is then represented by the number of active users at time $t$ $u_m(t)$, multiplied by $R_m$. We can define a Markov chain over $u_m\in\{0,\ldots,U_m\}$, with the following transition probabilities:
\begin{equation}
\begin{aligned}
P(u_m(t+1)=v | u_m(t) = u)=\sum_{\mathclap{w=\max(0,u+v-U_m)}}^{\mathclap{\min(u,v)}}({O}^{(m)}_{11})^w({O}^{(m)}_{10})^{u-w}\\
\times\binom{u}{w}\binom{U_m-u}{v-w}({O}^{(m)}_{01})^{v-w}({O}^{(m)}_{00})^{U_m-u-v+w}.
\end{aligned}
\end{equation}
The expected traffic $G_m$ from slice $m$ can be computed as:
\begin{equation}
    \E{G_m}=\frac{O^{(m)}_{01}U_mR_m}{O^{(m)}_{01}+O^{(m)}_{10}}.
\end{equation}
On the other hand, the total channel capacity is simply given by:
\begin{equation}
    C=\frac{NL}{\tau}=7.68\text{~MB/s}.
\end{equation}
We also consider different policies for each slice's queue: in both queues, packets that find a full queue are \emph{rejected}, i.e., they do not enter the queue and are dropped immediately. However, in the second queue, packets still in the queue whose age is higher than the deadline are also dropped, as they do not contribute to the \gls{qos} of the slice and transmitting them would just waste resources.

\subsection{Learning Plane}

We can then define the \gls{drl} agent and learning plane optimization parameters. We used a \gls{dqn}~\cite{mnih2015human} for the agent, as the problem is simple enough not to require more advanced architectures. We also simplified the state and action space, limiting the possible resource allocation vectors to the following:
\begin{equation}
   \mb{a}_{t+1}=\mb{a}_t+\delta_t, 
\end{equation}
with $\delta_t\in\{(1,-1),(0,0),(-1,1)\}$. In other words, the change in the allocation is at most 1 resource with respect to the previous \gls{drl} step. The outputs of the \gls{dqn} correspond to the estimated long-term value of selecting each $\delta_t$, so the network only has 3 output values. The state was also simplified: for each slice $m\in\{1,2\}$, the input to the network is given by the following values:
\begin{itemize}
    \item The number $q_m\in\{0,\ldots,Q\}$ of packets in the queue;
    \item The minimum latency $T^{\min}_m$ for packets transmitted in the previous slot;
    \item The maximum latency $T^{\max}_m$ for packets transmitted in the previous slot;
    \item The average latency $T^{\text{avg}}_m$ for packets transmitted in the previous slot;
    \item The number $d_m$ of dropped or rejected packets in the previous slot;
    \item The current number $a_m$ of resources allocated to the slice.
\end{itemize}
The values for each queue are contained in the tuple $\mb{s}^{(m)}=(q_m,T^{\min}_m,T^{\max}_m,T^{\text{avg}}_m,d_m,a_m)$, to which we add another parameter: $\xi^{(2)}$, i.e., the number of packets in the queue for slice 2 that would miss their deadline if they are not transmitted in the next slot. We can define it as follows:
\begin{equation}
    \xi^{(2)}=\sum_{i=1}^{Q}f_2\left(\Delta_{q_2(i)}\right)-f_2\left(\Delta_{q_2(i)}+1\right).
\end{equation}
As the first slice does not have latency requirements, there are no equivalent parameters for it.
The input to the \gls{dqn} is then given by $\mb{s}^{(m)},\mb{s}^{(m)},\xi^{(m)}$, for a total of 13 values, which are normalized to fit in the range between 0 and 1. The full network architecture is given in Table~\ref{tab:agent}\footnote{The complete implementation of the DQN agent and dynamic resource allocation is available at~\href{https://github.com/slahmer97/costoflearning}{Github}}.

\setlength{\textfloatsep}{15pt}
\begin{table}[t]
    \centering
    \caption{\gls{dqn} architecture.}
    \footnotesize
    {\renewcommand{\arraystretch}{1.25}
    \begin{tabular}{@{}ccc@{}}
    \toprule
     \multicolumn{2}{c}{\textbf{Layer size}}           &\multirow{2}{*}{\textbf{Activation function}}\\
    Input & Output &\\
     \midrule
     13 & 64 & ReLU\\
     64 & 32 & ReLU\\
     32 & 3 & Linear\\
     \bottomrule
    \end{tabular}}
    \label{tab:agent}
\end{table}

We then consider the learning plane optimization. First, we set the size of the experience sample queue $E=1500$, and implement an early rejection policy. When a sample is generated, its rejection probability is equal to $\frac{e}{E}$, i.e., to the current pressure on the queue. Consequently, samples that find a full queue are always rejected, but sometimes samples that could fit in the queue are dropped in favor of new experiences, avoiding too many correlated samples filling the queue.

The probability of selecting a slot as a learning slot decays linearly, starting from an initial value $\rho_0$ and gradually decaying to a value $\rho_f$. The decay is applied every $H$ steps, after which the learning rate decreases by a constant value $\sigma$:
\begin{equation}
    \rho(t)=\max\left(\rho_f,\rho_0-\left\lfloor\frac{t}{H}\right\rfloor\sigma\right).
\end{equation}

Finally, we consider the greedy allocation in the learning slots. As the first slice has no latency requirements, we consider allocating resources to it greedily only when the number of packets in the queue is higher than a threshold $\chi_1$: in this way, we avoid packet rejections, but also leave more resources for learning plane and latency-sensitive packets.

We can then define the following estimated rewards:
\begin{align}
    \hat{R}_1(\mb{s},\mb{z})&=\min(0,z_1-\min(q_1-\chi_1,N));\\
    \hat{R}_2(\mb{s},\mb{z})&=\min(0,z_2-\min(\xi_2,N));\\
    S(\mb{z},e)&=\min(e,N)-\left(N-\sum_{m=1}^2 z_m\right).
\end{align}
The minimum operation ensures that resources will not be allocated to a slice once the queue pressure is below the limit $\xi_1$ or all packets with a close deadline are served, respectively.
We can define the following problem:
\begin{equation}
\begin{aligned}
    \mb{z}^*(\mb{s},\mb{z},e)=\argmax_{\mb{z}\in\mc{Z}} S(\mb{z},e)+\sum_{m=1}^2\hat{R}_m(\mb{s},\mb{z}).
\end{aligned}
\end{equation}
As the problem can easily be converted to an integer linear problem, we can easily solve it through iterative methods by assigning each resource.

\section{Simulation Results}\label{sec:results}

\begin{figure}[t!]
\centerline{\includegraphics[width=3.3in]{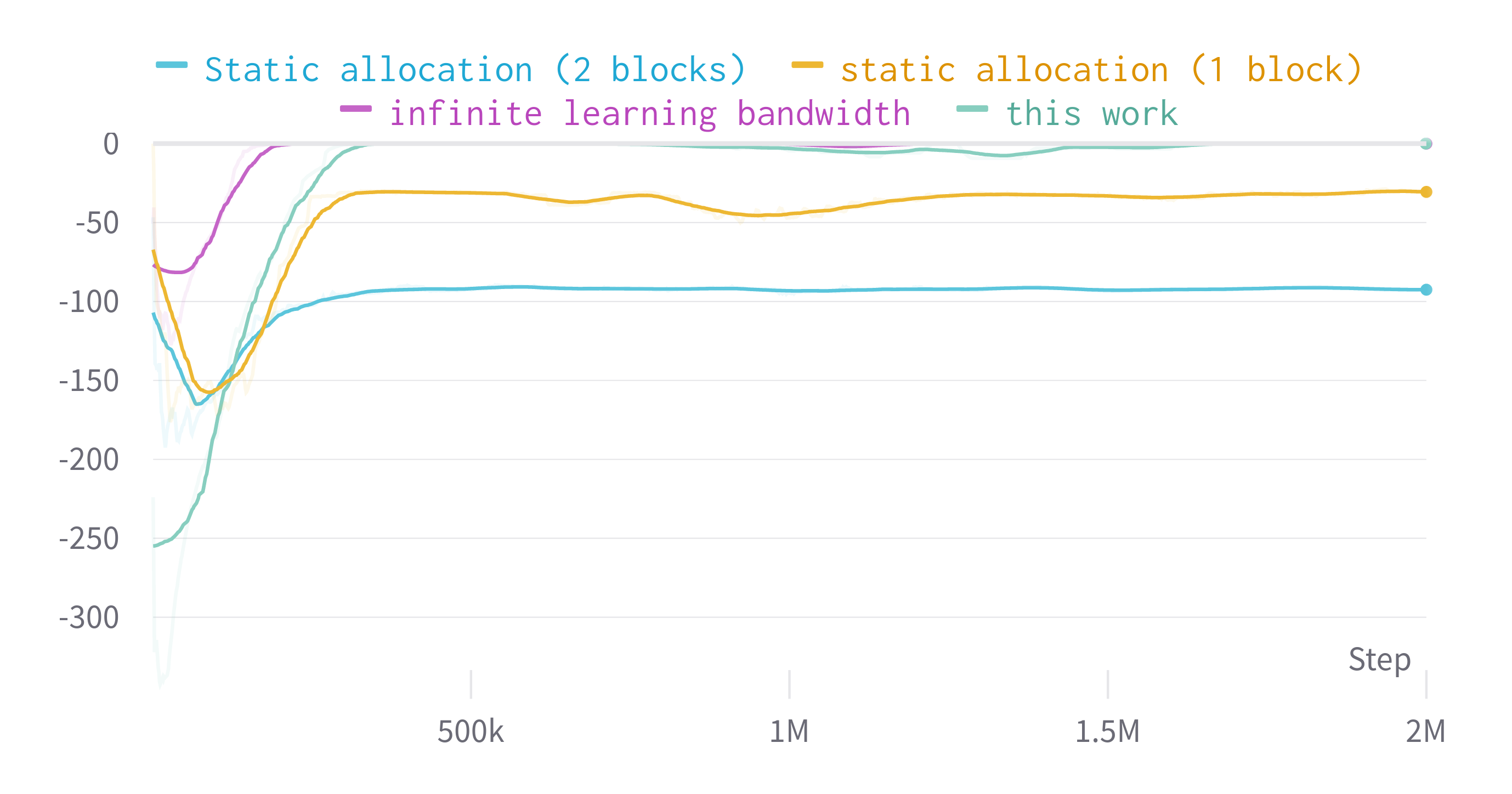}}
\caption{Average cumulative reward per second.}\vspace{-0.3cm}
\label{fig:averageReward}
\end{figure}

We can then measure the performance of the dynamic learning plane resource allocation, running the \gls{drl} agent in the slicing use case for 2000 seconds. We define two benchmarks against which we compare our dynamic solution:
\begin{itemize}
    \item \emph{Out-of-band} learning plane: in this ideal scenario, training data is transmitted over a side channel with infinite capacity. This corresponds to the common assumption in the literature of free training;
    \item \emph{Static} allocation: in this case, we set aside a fixed number of resources for the training process in each time slot. This is equivalent to the solution in our previous work~\cite{mason2022no}, which considered the cost of learning but only proposed static policies.
\end{itemize}
We run two versions of the static allocation, which give 1 and 2 resources per time slot to the learning plane, respectively.

The overall reward over the training process is shown in Fig.~\ref{fig:averageReward}: as the figure shows, the dynamic allocation initially performs worse than static allocation, but manages to converge slightly faster and maintain close to ideal performance after approximately $4\times10^5$ samples, corresponding to slightly more than 10 minutes of real time, while the static algorithms have a lower reward at convergence. This is also due to the fact that, unlike in our previous work, the resources allocated to the learning plane change the nature of the problem, so that resources allocated to the learning plane cannot just be transferred to the data plane: as the \gls{dqn} agent trained with $N'<N$ resources, it will perform suboptimally if it is asked to allocate $N$ resources.

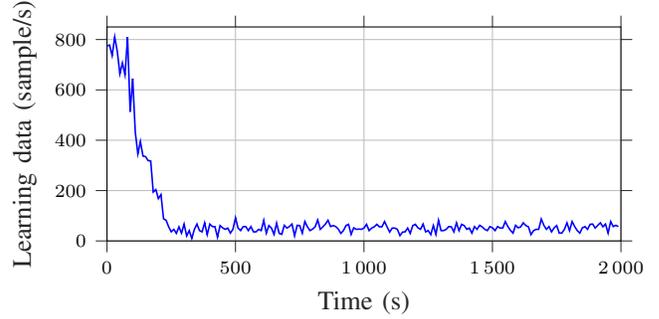
\begin{figure}[t]
\centering
\input{tikz-figures/averageForwardedExperiences.tex}
\caption{Average number of forwarded learning experiences per second.}\vspace{-0.3cm}
\label{fig:forwardedExperiences}
\end{figure}

\begin{figure}[t]
\centering
\input{tikz-figures/overheadPerGreedyAction.tex}
\caption{Empirical CDF of the reward loss during learning slots.}\vspace{-0.3cm}
\label{fig:greedyActionPerfOverhead}
\end{figure}
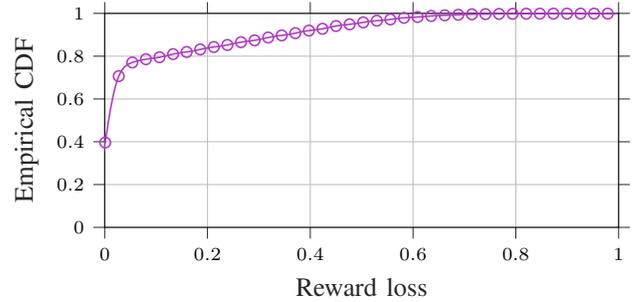

\begin{figure*}[t!]
    \centering
\subfloat[%
  Slice 1 (rejected packets). \label{fig:drop_1}%
]{\input{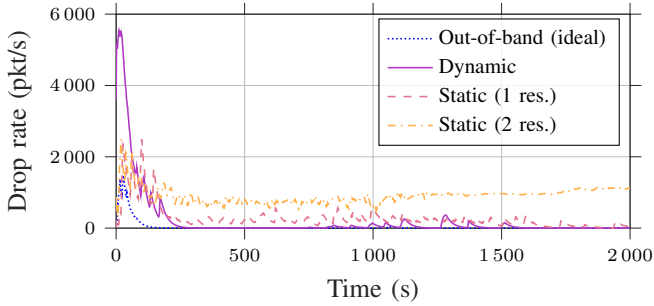}}
\subfloat[%
  Slice 2 (rejected and dropped packets). \label{fig:drop_2}%
]{\input{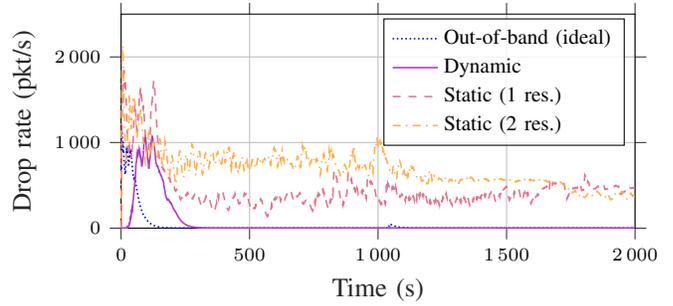}}
    \caption{Drop and rejection rates of each slice queue over time.}\vspace{-0.3cm}
    \label{fig:queue_drop}
\end{figure*}

We can also consider the effect of learning slots on the learning and data planes: Fig.~\ref{fig:forwardedExperiences} shows how many experience samples are forwarded to the Cloud during the training process. Following the linear decay of $\rho(t)$, the number of new experience samples transmitted for training is initially very high, but decreases over about 300 seconds to reach the minimum, which is between 40 and 50 samples per second. This rate is high enough to guarantee that changes in the environment statistics are captured, but does not impact the final performance, as we discussed above. Furthermore, we can analyze the impact of learning slots on the instantaneous reward by looking at the empirical \gls{cdf} of the reward penalty from using the greedy allocation, shown in Fig.~\ref{fig:greedy}: the reward loss is 0 in 40\% of cases, and below 0.1 in 80\% of cases. This means that the greedy allocation can still guarantee good performance in most cases, and as such, is a robust strategy for the learning slots. The static allocation with only 1 resource per time slot works better than with 2, even though its convergence is slightly slower, but still ends up dropping or rejecting hundreds of packets per second: slice 2 tends to suffer more, as it has more stringent requirements which need to be optimized better. On the other hand, our dynamic resource allocation manages to allocate resources as well as the ideal system, after the initial training period.

\begin{figure}[t]
\centering
\input{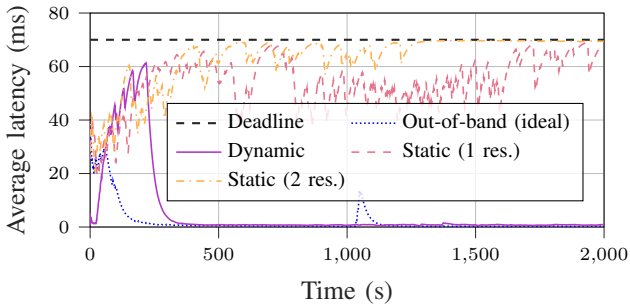}
\caption{Average queuing latency in slice 2.}\vspace{-0.3cm}
\label{fig:queue2AverageLatency}
\end{figure}

We can further analyze the performance of slice 2 by considering the average latency experienced by packets, as shown in Fig.~\ref{fig:queue2AverageLatency}: the out-of-band solution quickly converges to immediately serving all packets from the slice, ensuring the lowest possible risk of violating the hard timing requirement. The dynamic resource allocation needs more time, and approaches the limit during training as several resources are taken up by the learning plane, but quickly reduces the latency to the same level as the ideal policy, with only minor spikes that remain below 5~ms. On the other hand, the static strategies have a latency that is close to the limit, as the limited resources available to the data plane require much riskier choices to avoid having too much of an impact on the other slice. The dynamic allocation manages to achieve better performance for both slices, even during training, only suffering an initial performance drop in the first minute of training.

\section{Conclusions and Future Directions}\label{sec:conc}
In this work, we have designed a dynamic resource allocation policy which can mediate between the learning and data planes, controlling the trade-off between effectiveness and efficiency of \gls{drl} models. Unlike most works in the learning-based networking literature, we specifically consider the cost of learning, i.e., the resources required by the training process of a \gls{drl} agent, and shown that our dynamic policy can outperform static schemes and, after a short transition phase, match the performance of an ideal system with an out-of-band learning plane.

Possible extensions of the work include a wider deployment with more resources and a more difficult problem, with multiple slices and more stringent requirements, such as those for \gls{urllc}. Another interesting theoretical question is the interplay between cost of learning and active learning: when resources in the learning plane are scarce, transmitting the most valuable samples can significantly accelerate training. In \gls{drl}, this also has an important effect on the trade-off between exploration and exploitation, which we plan to study in a future work.

\bibliographystyle{IEEEtran}
\bibliography{biblio.bib}

\end{document}

%% file: tikz-figures/averageForwardedExperiences.tex
\begin{tikzpicture}

\definecolor{darkslategray38}{RGB}{38,38,38}
\pgfplotsset{every tick label/.append style={font=\scriptsize}}
\pgfplotsset{scaled x ticks=false}
\pgfkeys{/pgf/number format/.cd,1000 sep={\,}}

\begin{axis}[
width=\fwidth,
height=\fheight,
tick align=outside,
xlabel=\textcolor{darkslategray38}{Time (s)},
xmajorgrids,
xmajorticks=true,
xmin=0, xmax=2000,
xtick style={color=darkslategray38},
ylabel=\textcolor{darkslategray38}{Learning data (sample/s)},
ymajorgrids,
ymajorticks=true,
ymin=0, ymax=850,
ytick style={color=darkslategray38}
]
\addplot [color0,semithick]
table {%
0 775
10 778
20 736
30 810
40 753
50 664
60 707
70 660
80 810
90 513
100 645
110 432
120 346
130 395
140 337
150 335
160 319
170 318
180 194
190 204
200 168
210 184
220 88
230 82
240 56
250 36
260 46
270 30
280 56
290 31
300 67
310 20
320 41
330 10
340 46
350 67
360 41
370 36
380 72
390 25
400 67
410 57
420 57
430 15
440 61
450 52
460 46
470 51
480 31
490 46
500 92
510 51
520 41
530 57
540 57
550 41
560 57
570 36
580 36
590 46
600 41
610 82
620 31
630 61
640 51
650 26
660 71
670 30
680 26
690 62
700 51
710 57
720 67
730 20
740 62
750 61
760 31
770 77
780 56
790 41
800 47
810 56
820 83
830 46
840 56
850 67
860 82
870 57
880 61
890 57
900 47
910 30
920 41
930 61
940 66
950 26
960 52
970 46
980 47
990 46
1000 51
1010 66
1020 41
1030 51
1040 56
1050 66
1060 56
1070 56
1080 77
1090 56
1100 35
1110 51
1120 51
1130 46
1140 21
1150 35
1160 36
1170 51
1180 30
1190 61
1200 66
1210 52
1220 46
1230 67
1240 36
1250 41
1260 25
1270 61
1280 26
1290 82
1300 41
1310 41
1320 46
1330 57
1340 56
1350 36
1360 72
1370 41
1380 66
1390 61
1400 46
1410 51
1420 31
1430 57
1440 46
1450 62
1460 56
1470 46
1480 41
1490 57
1500 52
1510 41
1520 57
1530 51
1540 51
1550 77
1560 57
1570 42
1580 31
1590 62
1600 52
1610 56
1620 72
1630 61
1640 56
1650 25
1660 41
1670 46
1680 46
1690 87
1700 62
1710 46
1720 57
1730 36
1740 52
1750 61
1760 31
1770 31
1780 46
1790 72
1800 41
1810 77
1820 57
1830 41
1840 31
1850 52
1860 57
1870 46
1880 66
1890 67
1900 51
1910 62
1920 72
1930 56
1940 67
1950 31
1960 77
1970 57
1980 62
1990 57
};
\end{axis}

\end{tikzpicture}

%% file: tikz-figures/overheadPerGreedyAction.tex
\begin{tikzpicture}

\definecolor{darkslategray38}{RGB}{38,38,38}
\pgfplotsset{every tick label/.append style={font=\scriptsize}}

\begin{axis}[
width=\fwidth,
height=\fheight,
tick align=outside,
xlabel=\textcolor{darkslategray38}{Reward loss},
xmajorgrids,
xmajorticks=true,
xmin=0, xmax=1,
xtick style={color=darkslategray38},
ylabel=\textcolor{darkslategray38}{Empirical CDF},
ymajorgrids,
ymajorticks=true,
ymin=0, ymax=1,
ytick style={color=darkslategray38}
]
\addplot [semithick, color1, mark=o,mark repeat=5]
table {%
-0.0571733563488781 0
-0.0518876944760251 0.00144288538966454
-0.046602032603172 0.0044985191209931
-0.041316370730319 0.0104932813528343
-0.0360307088574659 0.0213887255255372
-0.0307450469846129 0.0397338753587905
-0.0254593851117598 0.0683494165182956
-0.0201737232389068 0.109700626084916
-0.0148880613660537 0.165058794858702
-0.00960239949320068 0.233715907535937
-0.00431673762034763 0.312603053593691
0.000968924252505415 0.396579775283692
0.00625458612535846 0.479408150025411
0.0115402479982115 0.555117885707128
0.0168259098710646 0.619280089974833
0.0221115717439176 0.669751448217996
0.0273972336167706 0.706699780197422
0.0326828954896237 0.732028106602767
0.0379685573624767 0.748514185281699
0.0432542192353298 0.759001377105755
0.0485398811081828 0.76585473216322
0.0538255429810359 0.770733859974999
0.0591112048538889 0.774616327848509
0.064396866726742 0.77796108723705
0.069682528599595 0.780912920860594
0.0749681904724481 0.783483321142658
0.0802538523453011 0.785676997010047
0.0855395142181541 0.787557490224697
0.0908251760910072 0.789259989352021
0.0961108379638602 0.790966565612754
0.101396499836713 0.792861058206875
0.106682161709566 0.795080039700368
0.111967823582419 0.797674382846037
0.117253485455272 0.800592976123681
0.122539147328125 0.803694843132581
0.127824809200979 0.806787770031562
0.133110471073832 0.809682516381758
0.138396132946685 0.812245769945883
0.143681794819538 0.814435447281634
0.148967456692391 0.816308629357597
0.154253118565244 0.818001863454445
0.159538780438097 0.819691451088298
0.16482444231095 0.821545359326133
0.170110104183803 0.823678898131558
0.175395766056656 0.826124744681026
0.180681427929509 0.828824908445234
0.185967089802362 0.831647421915957
0.191252751675215 0.834423919305766
0.196538413548068 0.836997571158371
0.201824075420921 0.839267085328363
0.207109737293774 0.841213844883725
0.212395399166627 0.842905249926334
0.21768106103948 0.844475009288227
0.222966722912333 0.846087172155026
0.228252384785186 0.847893569244603
0.233538046658039 0.849994663853567
0.238823708530893 0.852412860506925
0.244109370403746 0.855085343659243
0.249395032276599 0.857879694419262
0.254680694149452 0.860629491308779
0.259966356022305 0.863180409180042
0.265252017895158 0.865433135086866
0.270537679768011 0.867370252847304
0.275823341640864 0.869059913314795
0.281109003513717 0.870636677904951
0.28639466538657 0.872265944092688
0.291680327259423 0.874101254564087
0.296965989132276 0.876244295015378
0.302251651005129 0.878716852186755
0.307537312877982 0.88145250116386
0.312822974750835 0.884312087600395
0.318108636623688 0.887120568621174
0.323394298496541 0.88971529497413
0.328679960369394 0.891991078936469
0.333965622242247 0.893928316544762
0.3392512841151 0.895596672923172
0.344536945987953 0.897134965382149
0.349822607860806 0.898714091375002
0.355108269733659 0.900492668184641
0.360393931606513 0.902575534638955
0.365679593479366 0.904984858015883
0.370965255352219 0.90765209660525
0.376250917225072 0.910434980041461
0.381536579097925 0.913156505380501
0.386822240970778 0.915655020731107
0.392107902843631 0.917829900059272
0.397393564716484 0.919668899718288
0.402679226589337 0.921250147492913
0.40796488846219 0.922720002926915
0.413250550335043 0.924254038073619
0.418536212207896 0.926011069569153
0.423821874080749 0.928090810268054
0.429107535953602 0.930505479592442
0.434393197826455 0.933173943879158
0.439678859699308 0.93594203255132
0.444964521572161 0.938624619341582
0.450250183445014 0.941056922405724
0.455535845317867 0.943138693374678
0.46082150719072 0.944857919213238
0.466107169063574 0.946288675997372
0.471392830936427 0.947566428053935
0.47667849280928 0.948849586175361
0.481964154682133 0.95027773827611
0.487249816554986 0.951936237194189
0.492535478427839 0.953835277518765
0.497821140300692 0.955909043646995
0.503106802173545 0.958035943397097
0.508392464046398 0.960074608550786
0.513678125919251 0.961904699744989
0.518963787792104 0.963459755895067
0.524249449664957 0.964742639839192
0.52953511153781 0.965820740108878
0.534820773410663 0.966804506068546
0.540106435283516 0.967816585504483
0.545392097156369 0.968959645804262
0.550677759029222 0.970290259251935
0.555963420902075 0.971804967874159
0.561249082774928 0.973442433491305
0.566534744647781 0.975101677610895
0.571820406520634 0.976671345554765
0.577106068393487 0.978060874092268
0.582391730266341 0.97922378285752
0.587677392139194 0.980166628183141
0.592963054012047 0.980942635648805
0.5982487158849 0.981633827876428
0.603534377757753 0.982327787561768
0.608820039630606 0.983095176454309
0.614105701503459 0.983972995012853
0.619391363376312 0.984957151030874
0.624677025249165 0.986006061721263
0.629962687122018 0.987054338760993
0.635248348994871 0.988032513604174
0.640534010867724 0.988886634746621
0.645819672740577 0.989591743057248
0.65110533461343 0.990155789116183
0.656390996486283 0.990614141034528
0.661676658359136 0.991017654572665
0.666962320231989 0.991418376103754
0.672247982104842 0.991856588267383
0.677533643977695 0.992351940016883
0.682819305850548 0.992900359618469
0.688104967723401 0.993477261596516
0.693390629596255 0.994046033297407
0.698676291469108 0.994569198769335
0.703961953341961 0.995018761929256
0.709247615214814 0.995382654969247
0.714533277087667 0.99566585158058
0.71981893896052 0.995886700618985
0.725104600833373 0.996070422948443
0.730390262706226 0.996242053118394
0.735675924579079 0.996420639575633
0.740961586451932 0.996615778091671
0.746247248324785 0.996826893612378
0.751532910197638 0.997045125926142
0.756818572070491 0.997257122595148
0.762104233943344 0.997449560685321
0.767389895816197 0.997613038931736
0.77267555768905 0.997744278343392
0.777961219561903 0.997846247948355
0.783246881434756 0.997926561305051
0.788532543307609 0.997994937513687
0.793818205180462 0.998060577405941
0.799103867053315 0.998130096927859
0.804389528926168 0.998206383149806
0.809675190799022 0.99828849935439
0.814960852671875 0.998372553624744
0.820246514544728 0.998453231042778
0.825532176417581 0.998525514266421
0.830817838290434 0.99858607636764
0.836103500163287 0.99863397716787
0.84138916203614 0.998670572886934
0.846674823908993 0.998698820685732
0.851960485781846 0.998722304440732
0.857246147654699 0.998744299410286
0.862531809527552 0.998767091610822
0.867817471400405 0.998791651863005
0.873103133273258 0.998817673896528
0.878388795146111 0.998843916088112
0.883674457018964 0.998868725620984
0.888960118891817 0.99889058276384
0.89424578076467 0.998908507573744
0.899531442637523 0.99892223222902
0.904817104510376 0.998932135702063
0.910102766383229 0.998939018402106
0.915388428256082 0.998943827658952
0.920674090128936 0.998947426989389
0.925959752001788 0.998950456671958
0.931245413874642 0.998953289426072
0.936531075747495 0.998956059064983
0.941816737620348 0.998958731720134
0.947102399493201 0.998961190874585
0.952388061366054 0.998963312967752
0.957673723238907 0.998965018080067
0.96295938511176 0.998966289964085
0.968245046984613 0.998967169620777
0.973530708857466 0.998967733429822
0.978816370730319 0.998968068252627
0.984102032603172 0.998968252467877
0.989387694476025 0.998968346364127
0.994673356348878 0.998968390702125
};
\end{axis}

\end{tikzpicture}